\documentclass[a4paper]{article}
\usepackage{algorithm}
\usepackage{bbding}
\usepackage{algorithmic}
\usepackage{INTERSPEECH_v2,amssymb,amsmath,graphicx,tabularx,epstopdf,epsfig,booktabs,array}
\usepackage{multirow}
\usepackage{cite}
 \usepackage{algorithm}
\usepackage{algorithmic}        %

\usepackage{multirow}
\newcolumntype{I}{!{\vrule width 0.5pt}}
\newlength\savedwidth

\newlength\savewidth

\ninept
\def\reg{{\rm\ooalign{\hfil
     \raise.07ex\hbox{\scriptsize R}\hfil\crcr\mathhexbox20D}}}

\title{Integrated Speech Enhancement Method Based on Weighted Prediction Error and DNN for Dereverberation and Denoising}

\makeatother \name{{\em Hao Li\thanks{This research was partly supported by the China National Nature Science Foundation (No.61365006). \newline  \indent \quad $*$ Corresponding author. }, Xueliang Zhang$^*$, Hui Zhang, Guanglai Gao}}

\address{College of Computer Science, Inner Mongolia University, China}
\email{ lihao.0214@163.com   cszxl@imu.edu.cn   alzhu.san@163.com   csggl@imu.edu.cn}

\begin{document}
\maketitle

\begin{abstract}
Both reverberation and additive noises degrade the speech quality and intelligibility. the weighted prediction error (WPE) performs well on dereverberation but with limitations. First, The WPE doesn't consider the influence of the additive noise which degrades the performance of dereverberation. Second, it relies on a time-consuming iterative process, and there is no guarantee or a widely accepted criterion on its convergence. In this paper, we integrate deep neural network (DNN) into WPE for dereverberation and denoising. DNN is used to suppress the background noise to meet the noise-free assumption of WPE. Meanwhile, DNN is applied to directly predict spectral variance of the target speech to make the WPE work without iteration. The experimental results show that the proposed method has a significant improvement in speech quality and runs fast.

\end{abstract}
\noindent{\bf Index Terms}: Speech enhancement, DNN, WPE

\section{Introduction}
\label{sec:intro}
In real-world environments, speech signals are often contaminated by additive noises. In an enclosed space, such as a living room, the signals are also corrupted by their reflections from walls and other surfaces. Both the reverberation and additive noises degrade the audible quality of speech signal and the performance of automatic speech recognition (ASR)\cite{conf/icassp/VinyalsRP12 ,Gibak2009An, Demir2013Single}. Therefore, a speech enhancement processing is requested to deal with the reverberation and noises simultaneously.

Over the past decades, several single- and multi-channel dereverberation approaches have been proposed, which can be broadly categorized into acoustic channel equalization\cite{Cauchi2015Late ,conf/icassp/KodrasiJD16 ,Yoshioka2009Integrated}, spectral enhancement\cite{Emanu2009Late} and probabilistic model-based approaches\cite{conf/icassp/Habets05, journals/taslp/NakataniJYKDM08}. Acoustic channel equalization techniques remove the reverberation by reconstructing the room impulse responses (RIRs) between the acoustic source and the microphone. Although perfect dereverberation is achievable in theory, the performance of such methods heavily depends on the RIRs estimation, which is a tough problem in practice\cite{journals/taslp/MertinsMK10, journals/taslp/KodrasiGD13}. Most spectral enhancement methods, assuming that the early and later reflections are mutually independent, are derived from speech denoising methods. These methods always introduce disruptive speech distortion. Other speech dereverberation approaches are based on statistical acoustics model. One of the representative methods is WPE\cite{conf/icassp/NakataniYKMJ08, journals/taslp/NakataniYKMJ10} which models the reverberation with an autoregressive (AR) process and uses the Maximum likelihood (ML) estimation for dereverberation. To obtain the satisfactory results, the WPE relies on an iterative procedure to optimize the AR weights and desired speech spectral variance.

Although the WPE performs well, it has several limitations. First, since it is hard to satisfy the noise-free assumption, the performance of WPE is always influenced by background noise in real environments. Second, the original WPE employs an iterative procedure which is time consuming. Third, there is no guarantee on the convergence of the prediction weights and the performance may reduce when more iterations are applied\cite{journals/taslp/NakataniYKMJ10}. Therefore, several studies (e.g in\cite{Parchami2013Speech}) focus on replacing the iterative procedure.


Recently, deep neural networks (DNNs) show their power on the speech enhancement\cite{journals/taslp/ZhangW16a, Han2015Learning, conf/interspeech/LiNZZ16, Zhang2016A}. The DNN-based methods usually predict the magnitude spectrogram of interest signal or a mask to remove the undesired parts\cite{Wang2014On, Xu2014An, conf/icassp/NarayananW13a}. Although, the DNN can build the non-linear relationships between the mixture and the target by training on large amounts of data, it is totally data-driven without considering the speech signal processing theory. In this paper, the DNN and WPE are combined. The DNN is applied to remove the background noise and estimate the desired speech spectral variance directly, which can  reduce the influence of background noise to WPE and make it work without iteration.

The rest of this paper is organized as follows. In Section 2, the problem formulation and the WPE algorithm are introduced. Section 3 presents the proposed model and algorithm in detail. Experiments and evaluations are given in Section 4. Finally, Section 5 provides the conclusions.

\section{WPE METHOD}
\label{sec:Related Works}
In this section, a brief description of the WPE method from a statistical viewpoint is presented. A scenario where a single speech source is captured by $M$ microphones is considered. In the STFT domain, $s_{n,k}$ denotes the clean speech signal with time frame index $n\in\{1,...,N\}$ and frequency bin index $k \in \{1,...,K\}$. The speech signal at the $m$-th $\left(m\in\{1,...,M\}\right)$ microphone, $x^m_{n,k}$, can be modeled as
\begin{equation}
x^m_{n,k} = \sum_{l=0}^{L_h-1} \left (  h^m_{l,k}\right )^*s_{n-l,k} + e^m_{n,k}, \\
\label{eq1}
\end{equation}
where $h^m_{l,k}$ is an approximation of the acoustic transfer function (ATF) between the speech source and the $m$-th microphone with length of $L_h$, and $\left ( \cdot \right )^*$ denotes the complex conjugate operator. The additive term $e^m_{n,k}$ represents the sum of modeling errors and background noise. In WPE method, the $e_{n,k}^{m} (\forall n,k,m)$ is always assumed to 0. The convolutive model in (\ref{eq1}) is often rewritten as

\begin{equation}
x^m_{n,k} = \sum_{l=0}^{D-1} \left (  h^m_{l,k}\right )^*s_{n-l,k} + \sum_{l=D}^{L_h-1} \left (  h^m_{l,k}\right )^*s_{n-l,k}, \\
\label{eq2}
\end{equation}
where the $\sum_{l=0}^{D-1} \left (  h^m_{l,k}\right )^*s_{n-l,k}$ indicates the sum of the direct signal and the early reflection at the $m$-th microphone. And $\sum_{l=D}^{L_h-1} \left (  h^m_{l,k}\right )^*s_{n-l,k}$ denotes the later reflection. $D$ corresponds to the duration of the early reflections. To simplify the expression, we rewritten (\ref{eq2}) as
\begin{equation}
x^m_{n,k} = d^m_{n,k} + \sum_{l=D}^{L_h-1} \left (  h^m_{l,k}\right )^*s_{n-l,k}. \\
\label{eq3}
\end{equation}
Because the early reflections signal can actually improve speech intelligibility\cite{Cohen2010Speech}, most dereverberation methods aim to reconstruct $d^m_{n,k}$ as the desired speech. By replacing the convolutive model in (\ref{eq3}) with an AR model, the signal observed at the first microphone ($m=1$) can be rewritten in the well-known multi-channel linear prediction (MCLP) form
\begin{equation}
x^1_{n,k} = d^1_{n,k} + \sum_{m=0}^{M} \left (  \textbf{g}^m_{k}\right )^H \textbf{x}^m_{n-D,k} \\
\label{eq4}
\end{equation}
where $d^1_{n,k}$ is the desired signal, and $(\cdot)^H$ denotes the conjugate transposition operator. The vector $\textbf{g}^m_k \in \mathbb{C}^{L_k}$ is the regression vector of order $L_k$ for the $m$-th channel. $\textbf{x}^m_{n,k}$ and $\textbf{g}^m_k$ are defined as
\[
\textbf{x}^m_{n-D,k} = \left [ x^m_{n-D,k}, x^m_{n-D-1,k} , ..., x^m_{n-D-(L_k-1),k}  \right ]^T,  \\
\]
\begin{equation}
\begin{split}
\textbf{g}^m_{k} = \left [ g^m_{0,k} , g^m_{1,k} , ..., g^m_{L_k-1,k} \right ]^T,
\label{eq5}
\end{split}
\end{equation}
where $(\cdot)^T$ denotes the transposition operator. The MCLP model (\ref{eq4}) can be written in a compact form using the multi-channel regression vector $\textbf{g}_k \in \mathbb{C}^{ML_k}$
\begin{equation}
x^1_{n,k} = d_{n,k} + \textbf{g}^H_k \textbf{x}_{n-D, k} \\
\label{eq6}
\end{equation}
where
\[
d_{n,k}\equiv d^1_{n,k},  \\
\]
\[
\textbf{g}_k=\left[(\textbf{g}^1_{k})^T,...,(\textbf{g}^M_{k})^T \right]^T,  \\
\]
\begin{equation}
\begin{split}
\textbf{x}_{n, k}=\left[(\textbf{x}^1_{n,k})^T,...,(\textbf{x}^M_{n,k})^T\right]^T.
\label{eq7}
\end{split}
\end{equation}
From (\ref{eq6}), the desired signal can be rewritten as
 \begin{equation}
d_{n,k} = x^1_{n,k} - \textbf{g}^H_k \textbf{x}_{n-D, k}. \\
\label{eq8}
\end{equation}
In WPE method, the desired signal in each frequency bin can be modeled as a circular complex Gaussian distribution with zero-mean and frequency-dependent variance $\sigma^2_{d_{n,k}}$. Assuming independence across time frames, by using the maximum likelihood (ML) estimation of the desired speech $d_{n,k}$ at each frequency, the joint distribution of the desired speech coefficients at frequency bin, $k$, is given by
 \begin{equation}
p \left( \textbf{d}_k \right ) = \prod^N_{n=1} \frac{1}{\pi \sigma^2_{d_{n,k}}} \mathbf{\exp} \left (  - \frac{\left | d_{n,k} \right |^2}{\sigma^2_{d_{n,k}}} \right ), \\
\label{eq9}
\end{equation}
where $\sigma^2_{d_{n,k}}$ is the time-varying spectral variance of the desired speech. By inserting $d_{n,k}$ from (\ref{eq8}) into (\ref{eq9}) and taking the negative of logarithm of $p(\textbf{d}_k)$ in (\ref{eq9}), the objective function can be written as
 \begin{equation}
 \begin{split}
\mathcal{J}(\Theta_k) &= -\log p(\textbf{d}_k|\Theta_k) \\& = \sum_{n=1}^N (\log \sigma^2_{d_{n,k}} + \frac{|x^1_{n,k}-\textbf{g}^H_k \textbf{x}_{n-D, k}|^2}{\sigma^2_{d_{n,k}}}), \\
\label{eq10}
\end{split}
\end{equation}
where constant terms are ignored. $\Theta_k=\{\textbf{g}_k, \sigma^2_{d_{1,k}}, \sigma^2_{d_{2,k}}, ... , \sigma^2_{d_{N,k}}\}$ are the unknown parameters for the $k$-th frequency bin. These parameters can be split into two groups: $\{g_k\}$, the AR weights, and  $\{\sigma^2_{d_{2,k}}, ... ,\sigma^2_{d_{N,k}}\}$, the spectral variance of the desired speech. A two-step algorithm to minimize the objective function is adopted by optimize the AR weights and desired speech spectral variance, alternatively. First, fix the $\textbf{g}_k$, $\{\sigma^2_{d_{1,k}}, \sigma^2_{d_{2,k}}, ... ,\sigma^2_{d_{N,k}}\}$ are adopted to minimize objective function. The estimated ${\sigma}^2_{d_{n,k}}$ can be obtained by

 \begin{equation}
 {\hat{\sigma}^2_{d_{n,k}}} = |{d_{n,k}}|^2, \quad n=1,2,...,N. \\
\label{eq11}
\end{equation}
It is the square of the desired speech spectrum of the first channel, see (\ref{eq8}).
Then, fix the $\{\sigma^2_{d_{1,k}}, \sigma^2_{d_{2,k}}, ... ,\sigma^2_{d_{N,k}}\}$, and $\textbf{g}_k$ is applied to minimize the objective function. The estimated $\textbf{g}_k$ can be obtained by

\begin{equation}
\hat{\textbf{g}}_k=\left(\sum_{n=1}^N \frac{\textbf{x}_{n-D,k} \textbf{x}^{H}_{n-D,k}}{\sigma^2_{d_{n,k}}}\right)^{-1} \sum_{n=1}^N \frac{\textbf{x}_{n-D,k} (x^1_{n,k})^*}{\sigma^2_{d_{n,k}}}. \\
\label{eq12}
\end{equation}
The estimated $\hat{\textbf{g}}_k$ is then used in (\ref{eq8}) to obtain $d_{n,k}$. The above two steps are repeated until some convergence criterions are satisfied or a maximum number of iteration is exceeded.

\section{PROPOSED METHOD}
As described in section 1, the WPE method has three limitations: the noiseless assumption, high time complexity of iteration and no guarantee on the convergence. In this paper, DNN is introduced to overcome these three limitations. First, we generate noiseless signals to meet the WPE assumption. Second, the iteration is removed to reduce the time complexity by estimating the desired speech spectral variance directly with DNN. In original WPE, the iteration is required by the alternative optimization of the AR weights and the desired speech spectral variance. We obtain the desired speech spectral variance directly from the DNN, and only need to get the AR weights by (\ref{eq12}). Therefore, the iteration is not needed any more. Without iterations, there is no need to consider the convergence problem.

\begin{figure}[!t]
\centering
\setlength{\abovecaptionskip}{-0.cm}
  \setlength{\belowcaptionskip}{-0.cm}
\includegraphics[width=\linewidth]{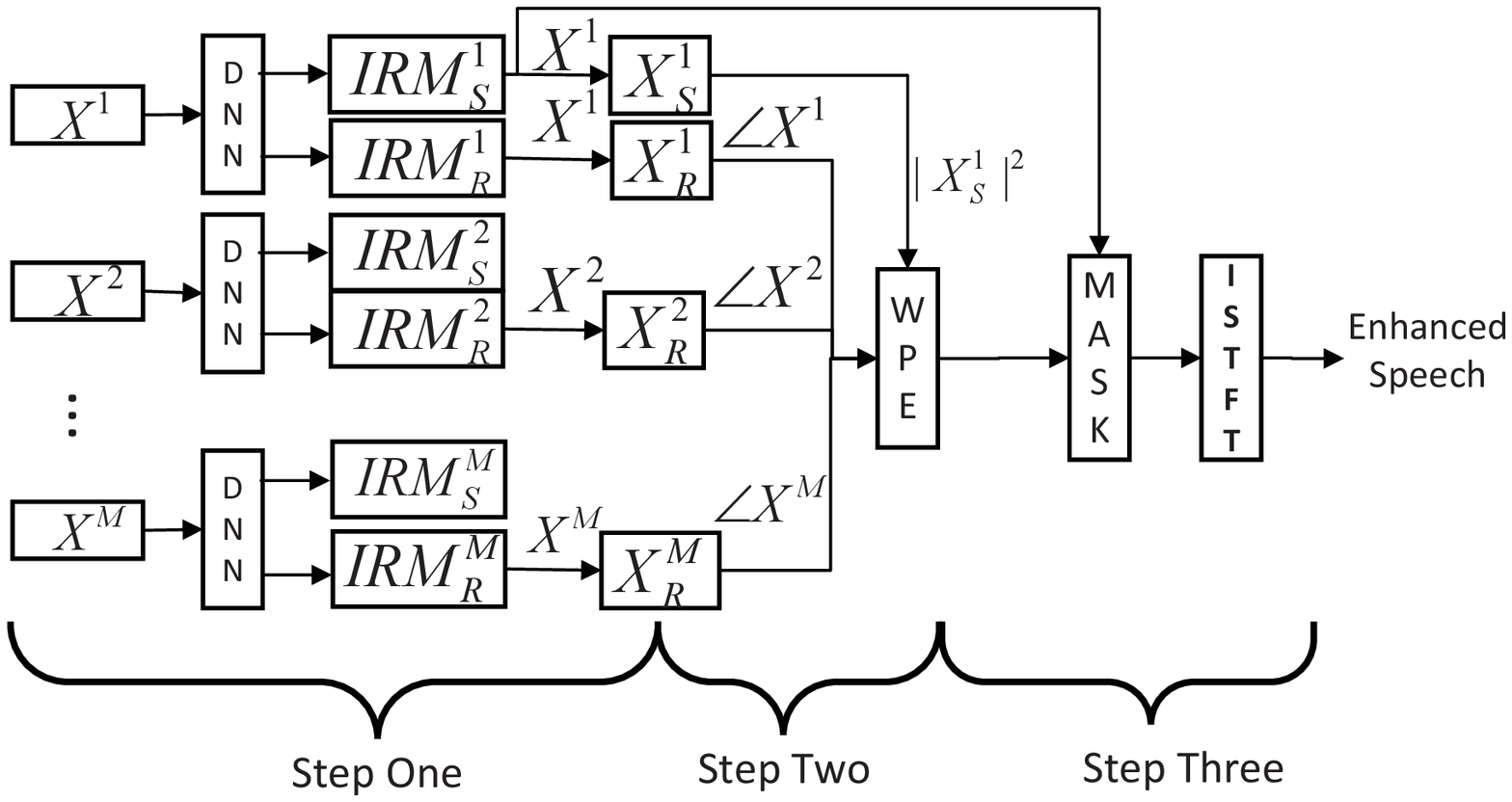}
\caption{\it Schematic diagram of proposed method.}
\label{fig:flow}
\end{figure}

We illustrate the proposed method in Figure~\ref{fig:flow}, and describe it in three steps: In the first step, the noise is removed and the desired speech spectral variance is estimated. Then, the output of WPE is obtained with the estimation of the AR weights. Finally, the residual noise is removed by applying the estimated mask received in the first step.

\subsection{Denoising and Variance Estimation}
\begin{figure}[!t]
\setlength{\abovecaptionskip}{0.0cm}
  \setlength{\belowcaptionskip}{-0.3cm}
\centering
\includegraphics[width=\linewidth]{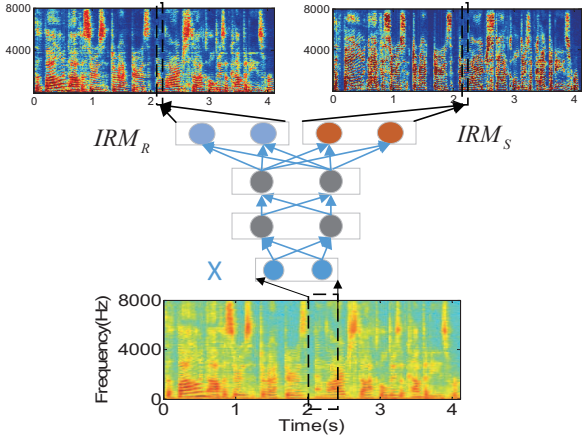}
\caption{\it Structure of the DNN. The inputs are the log spectra of the current frame and its neighboring frames. The outputs are $IRM_{R}$ and $IRM_{S}$ respectively. }
\label{fig:DNN}
\end{figure}

%
 A single channel DNN is used to predict the noiseless reverberation spectrum and the desired speech spectral variance estimation simultaneously. The structure of the DNN is shown in Figure~\ref{fig:DNN}. Eq.(\ref{eq11}) shows that the desired speech spectral variance $\sigma^2_{d_{n,k}}$ is the square of the desired speech spectrum of the first channel (refer to Eq.(\ref{eq8})).
We train a DNN to predict the noiseless reverberation spectrum $X_{R}^m$ and desired speech spectrum $X_{S}^m$ using ideal ratio mask $IRM^m_{R}$ and $IRM^m_{S}$\cite{journals/speech/SrinivasanRW06} as the targets which are defined as

 \begin{equation}
\begin{split}
IRM^m_{R} &= \min \{ \frac{X^m_{R}}{X^m+\epsilon}, 1  \} \\
IRM^m_{S} &= \min \{\frac{X^m_{S}}{X^m+\epsilon}, 1\}.
\label{eq13}
\end{split}
\end{equation}
After obtaining the estimated mask, the estimated noiseless reverberation spectrum $\hat{X}^m_{R}$ and the desired speech spectrum $\hat{X}^m_{S}$ are obtained by

 \begin{equation}
\begin{split}
\hat{X}^m_{R} &= X^m \otimes IRM^m_{R} \\
\hat{X}^m_{S} &= X^m \otimes IRM^m_{S} ,
\label{eq14}
\end{split}
\end{equation}
where $\otimes$ denotes an element-wise multiplication.

\subsection{Dereverberation}
After getting $\hat{X}_{R}$ and $\hat{X}^m_{S}$, the phase of mixture ($\angle{X}$) is used to reconstruct the noise-free phase spectrum $\tilde{X}_{R}$. The WPE is adopted to reconstruct the noise-free desired signal $\hat{d}_{n,k}$. $\tilde{X}_{R}$ and $|\hat{X}^1_{S}|^2$ are used as the input of WPE and $\sigma^2_{d_{n,k}}$ in Eq.(\ref{eq11}), respectively. The AR weights, $\textbf{g}_k$, can be obtained from (\ref{eq12}). Then the output of WPE, $\hat{d}_{n,k}$, can be calculated from (\ref{eq8}). The procedure is outlined in Algorithm (\ref{alg:Framwork_secondStep}).
       \begin{algorithm}[htb]
    \caption{ Dereverberation algorithm.}
    \label{alg:Framwork_secondStep}
    \begin{algorithmic} 
    \REQUIRE  
     $L_k$, $D$ \\
    $\textbf{initialization:}$ $\sigma^2_{d_{n,k}} \leftarrow |\hat{X}^1_{S}|^2$, $X \leftarrow \tilde{X}_{R}$\\
    \STATE $\textbf{A}_k \leftarrow \sum_{n=1}^N \frac{\textbf{x}_{n-D,k} \textbf{x}^{H}_{n-D,k}}{\sigma^2_{d_{n,k}}}$
    \STATE $\textbf{b}_k \leftarrow \sum_{n=1}^N \frac{\textbf{x}_{n-D,k} (x^1_{n,k})^*}{\sigma^2_{d_{n,k}}}$
    \STATE $\hat{\textbf{g}}_k \leftarrow \textbf{A}^{-1}_k \textbf{b}_k$
    \STATE $\hat{d}_{n,k} \leftarrow x^1_{n,k} - \hat{\textbf{g}}^H_k \textbf{x}_{n-D,k}$
    \end{algorithmic}
    \end{algorithm}

\subsection{Post processing}
The performance can be improved further by masking the $IRM^1_{S}$ to the output $\hat{d}_{n,k}$ to form the final enhanced spectrum. Finally the enhanced signal is converted from frequency domain to time domain using the inverse short time Fourier transform (ISTFT).

\section{Experiments}

\subsection{Experimental data and Metrics}
The official set of the 2014 REVERB Challenge2\cite{Kinoshita2016A} is used in the experiment. The data set consists of a training, a development and a (final) evaluation test set. The training and development sets are constructed using separated parts of the WSJCAM0\cite{wsjcam0} corpus convolved with 24 RIRs  and corrupted by various types of noises at 10 dB. The reverberation time (RT) of the 24 RIRs ranges roughly from 0.2 to 0.8 sec. Each RIR includes 8 channels. The test set includes simulated and ‘real world’ data sets. The simulated data is generated by convolving WSJCAM0 corpus with 6 RIRs (3 rooms, 2 types of distance between a speaker and a microphone array), and mixing with various types of noises at 10 dB. It should be mentioned that all the RIRs are recorded in real rooms and different for the training and simulated sets.

We randomly select 5000 sentences (only using the first channel) from the official training set to learn the DNN weights. 400 sentences are selected from the official simulated data set to evaluate the proposed method.

A fixed 50-ms frame size was used with $80\%$ overlap between frames. The discrete Fourier transform (DFT) is applied on each frame. The length of the DFT is 800.

The performance is evaluated with perceptual evaluation of speech quality (PESQ)\cite{conf/icassp/RixBHH01} and cepstral distance (CD)\cite{Furui1989Digital} which reflect the quality of the objective speech. The cepstral distance between two signals is defined as

%
%
%
%
 \begin{equation}
 CD = \frac{10}{\log10}\sqrt{(c_0 - \hat{c}_0)^2 + 2 \sum_{k=1}^{12}（(c_k - \hat{c}_k)^2）} \\
\label{eq15}
\end{equation}
where $c_k$ and $\hat{c}_k$ are the cepstral coefficients of the anechoic speech signal and the estimated desired signal, respectively.

\subsection{Experimental Setup and Parameter Selection }
To evaluate the performance, the single channel DNN in proposed method uses magnitude spectrogram with 5-frame context window (2 before and 2 after) as the input features.  The DNN has three hidden layers with 1024 rectified linear units (ReLUs)\cite{Glorot2010Deep} for each. Sigmoidal units are used in the output layer since the IRM is bounded between 0 and 1. The input is normalized to zero mean and unity variance over all feature vectors in the training set. The DNN weights are randomly initialized without pretrain. Root Mean Square Propagation (RMSprop)\cite{RMSprop} is utilized for optimization. The order of the regression vector $L_k$ and the prediction delay $D$ in WPE of the proposed method are set to 15 and 3 respectively.

Since the proposed algorithm (denoted as `Proposed') is independent on the number of microphones, we evaluate and compare the performances in single and multi-channel scenario, separately. To evaluate the single-channel enhancement performance, DNN-Han\cite{Han2015Learning} and original WPE are used to compare with the proposed method. DNN-Han use a DNN to learn the spectral mapping from reverberant and noisy signal to clean signal directly. In multi-channel speech enhancement experiment, we compare the proposed method with original WPE. In both single- and multi-channel experiments, the $L_k$ and the $D$ in original WPE are the same as the ones in the proposed method.


The original WPE cannot reduce the background noise. In order to compare fairly, we apply the $IRM^1_R$, obtained in the first step, to the output of the WPE for noise reduction. This is denoted as `WPE-mask'.

\subsection{Experimental results}

\begin{figure}[!t]
\centering
\setlength{\abovecaptionskip}{-0.cm}
  \setlength{\belowcaptionskip}{-0.cm}
\includegraphics[width=\linewidth]{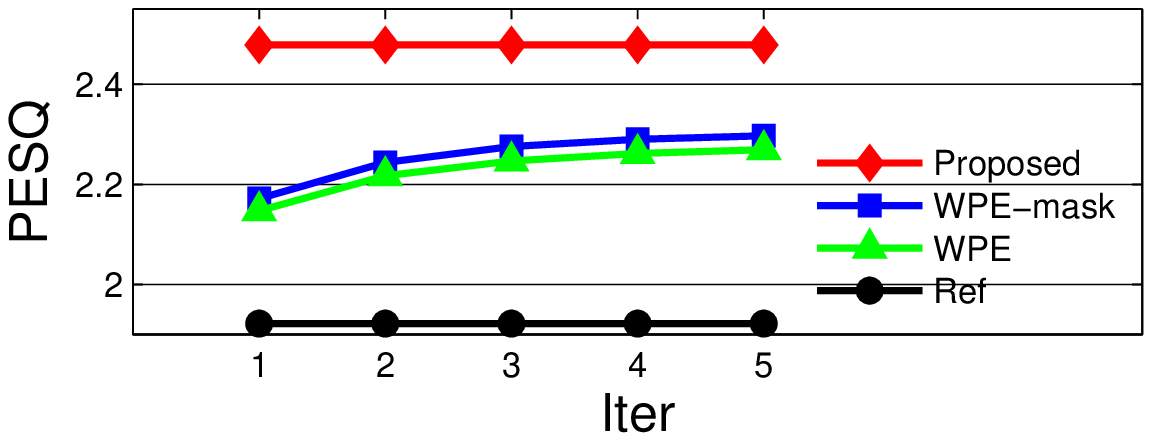}
\includegraphics[width=\linewidth]{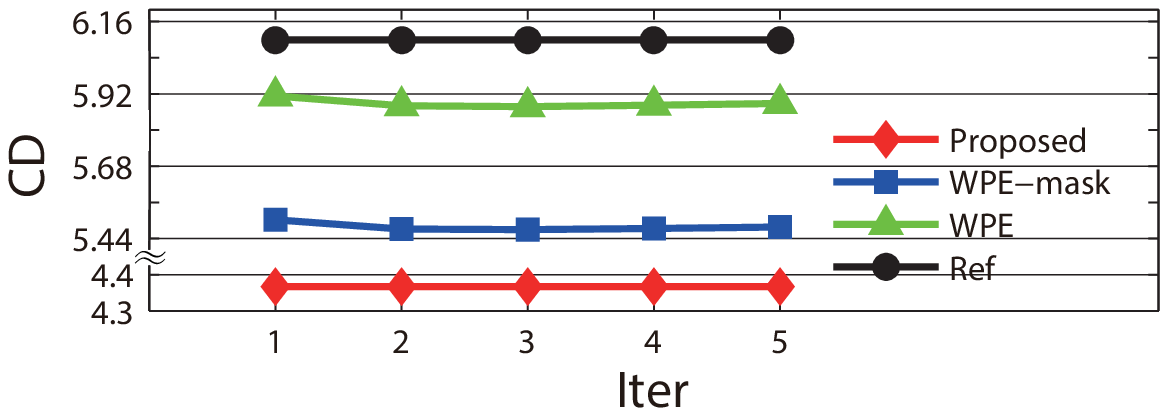}
\caption{\it PESQ and CD measures versus the numbers of iterations with various source algorithms for multi-channel speech enhancement. The reported values are obtained as the average of all test utterances.}
\label{fig:figure_multi}
\end{figure}

Figure~\ref{fig:figure_multi} shows a comparison of the proposed method and the WPE for multi-channel speech enhancement. ``Ref" indicates the noise and reverberation speech. The results show that the proposed method achieves best results on PESQ and CD. The WPE-mask outperforms the WPE because of noise reduction. However, it is still worse than the proposed method. It means that our method has better dereverberation ability due to the noise reduction and precise variance estimation by DNN.

\begin{figure}[!t]
\centering
\setlength{\abovecaptionskip}{-0.cm}
  \setlength{\belowcaptionskip}{-0.cm}
\includegraphics[width=\linewidth]{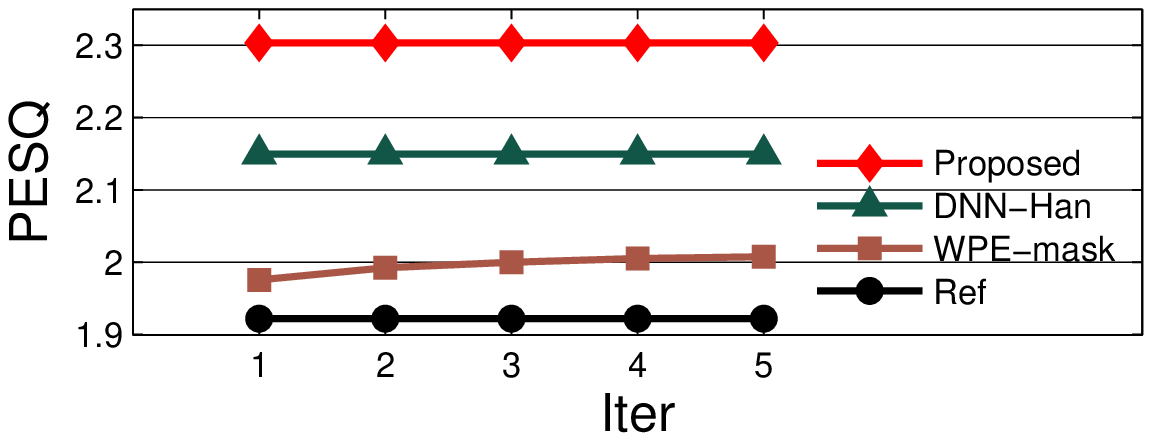}
\includegraphics[width=\linewidth]{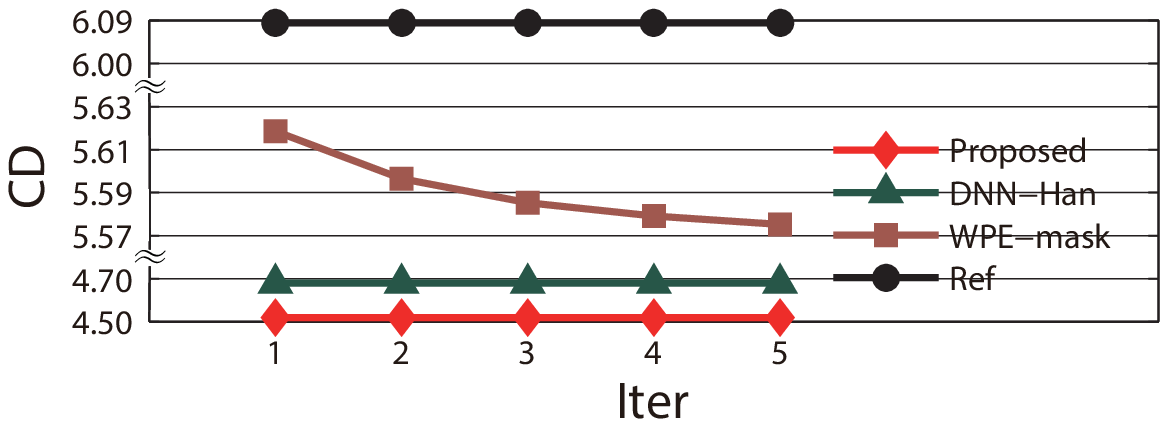}
\caption{\it PESQ and CD measures versus the numbers of iterations with various source algorithms for single-channel speech enhancement. The reported values are obtained as the average of all test utterances.}
\label{fig:figure_Single}
\end{figure}

The performance of single channel speech enhancement is shown in Figure~\ref{fig:figure_Single}. We can observe that the proposed method is still better than WPE-mask. In addition, the proposed method outperforms the DNN-Han. The possible reason is that the DNN-Han is a totally data-driven model which does not consider the statistical characteristics of speech.

\begin{figure}[!t]
\centering
\setlength{\abovecaptionskip}{-0.cm}
  \setlength{\belowcaptionskip}{-0.cm}
\includegraphics[width=\linewidth]{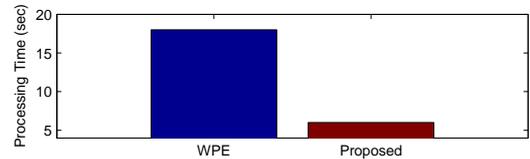}
\caption{\it Processing time required from a 10 secs. speech for WPE and proposed method. Intel(R) Xeon(R) CPU E5-2690 v2 @ 3.00GHz with RAM: 32G and Matlab environment.}
\label{fig:figure_times}
\end{figure}

The computational costs of the WPE and the proposed method are also evaluated. In WPE, the best performance is achieved with 5 iterations. The comparison result presented in Figure~\ref{fig:figure_times} shows that the computational effort has been considerably reduced by eliminating the iterative process.

\begin{figure}[!t]
\centering
\setlength{\abovecaptionskip}{-0.cm}
  \setlength{\belowcaptionskip}{-0.cm}
\includegraphics[width=\linewidth]{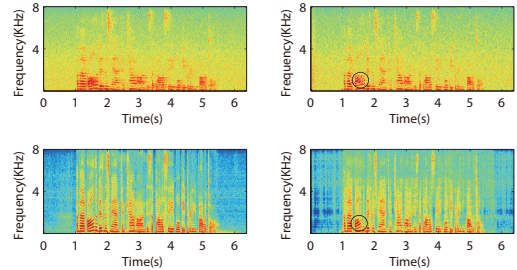}
\caption{\it Top left: The spectrogram of the mixture speech; Bottom left: Clean speech spectrogram; Top right: Speech separation using WPE; Bottom right: Speech separation using proposed method.}
\label{fig:figure_spec}
\end{figure}

Finally, an example of reconstructed spectrogram is shown in Figure~\ref{fig:figure_spec}. Obviously, our method has a better performance on denoising. As shown in the black ellipse, our method can also suppress more reverberations.

\section{Conclusion}
In this paper, a novel algorithm for speech denoising and dereverbration is proposed. The experimental results show that the proposed method outperforms the DNN and the conventional WPE. The proposed method takes advantage of the merits of the machine learning method: DNN and the conventional speech processing method: WPE. Using DNN, the influence of the background noise is reduced in the process of WPE and more accurate parameters are obtained directly. In addition, the proposed method can be used in single- or multi-channel speech enhancement, which is very flexible in practice. The proposed method can be easily extended to Recurrent Neural Networks (RNNs) and  Long Short Term Memory Networks (LSTMs) for better performance. We are also investigating this avenue.

\newpage
\eightpt

\bibliographystyle{}
\bibliography{refs}
\end{document}